\documentclass[10pt,twocolumn,letterpaper]{article}

\usepackage{cvpr}
\usepackage{times}
\usepackage{epsfig}
\usepackage{graphicx}
\usepackage{amsmath}
\usepackage{amssymb}
\usepackage{multirow}
\usepackage[hyphens]{url}

\usepackage[breaklinks=true,bookmarks=false,draft]{hyperref}

\cvprfinalcopy 

\def\cvprPaperID{****} 

\begin{document}

\title{Food for thought: Ethical considerations of user trust in computer vision}

\author{Kaylen J. Pfisterer$^{*,1,3}$, \text{Jennifer Boger}$^{2,3}$,\text{ and Alexander Wong}$^{1,3}$\\
$^{1}$ Vision and Image Processing Research Group, University of Waterloo, Waterloo, ON, Canada \\
$^{2}$ Intelligent Technologies for Wellness and Independent Living Lab, \\University of Waterloo, Waterloo, ON, Canada \\
$^{3}$ Schlegel-UW Research Institute for Aging, Waterloo, ON, Canada \\
$^{*}${\tt\small kpfisterer@uwaterloo.ca}
}

\maketitle

\begin{abstract}
In computer vision research, especially when novel applications of tools are developed, ethical implications around user perceptions of trust in the underlying technology should be considered and supported. Here, we describe an example of the incorporation of such considerations within the long-term care sector for tracking resident food and fluid intake. We highlight our recent user study conducted to develop a Goldilocks' quality horizontal prototype designed to support trust cues in which perceived trust in our horizontal prototype was higher than the existing system in place. We discuss the importance and need for user engagement as part of ongoing computer vision-driven technology development and describe several important factors related to trust that are relevant to developing decision-making tools.
\end{abstract}

\section{Introduction}

Ethical considerations around users' perceptions of trust in technology is a crucial but often neglected aspect when conducting computer vision research in areas that can have a major impact on societal well-being. This was highlighted in our recent work developing a computer vision application targeted at addressing malnutrition in older adults living in long-term care (LTC). In particular, our Automated Food Imaging and Nutrient Intake Tracking (AFINI-T) system is an end-to-end system driven by computer vision. For context, approximately 25\% of the population worldwide has some form of malnutrition (\eg , iron deficiency, zinc deficiency,  food-insecurity malnutrition)~\cite{rojer2016,kumssa2015,mcguire2015fao} and this risk increases with age ~\cite{brownie2006}. In LTC malnutrition prevalence in older adult residents is estimated to be 44\% ~\cite{keller2017}. Food and fluid intake tracking are mandated in LTC for those at-risk for malnutrition~\cite{OANHSS2011}, however existing methods rely on subjective, inaccurate, and typically retrospective methods~\cite{andrews2003,castellanos2002}. We aim to support improving quality of care, and thereby residents' quality of life, by harnessing computer vision to create an objective automation of this process in this historically subjective, paper-based application.

AFINI-T is a computer vision-based system comprised of four subsystems that perform cascading tasks: (1) \textit{Optical}: images are acquired using an integrated RGB-D camera; (2) \textit{Food segmentation}: discrimination between foods, the plate and the background; (3) \textit{Food volume estimation}: feature extraction to describe each food segment and classification of each food segment to inform the (4) \textit{Food intake and nutrient estimation}: draw inferences of the nutritional information for each food segment and summarize nutrient intake across a meal.

Others have made advances on aspects of this problem however, relatively few end-to-end systems have been developed~\cite{meyers2015,okamoto2016,pouladzadeh2016} and these approaches have focused on an individual tracking their own intake for the purpose of weight loss or maintenance. While supporting individual nutritional tracking is certainly needed with tripled obesity rates since 1975~\cite{WHO2018}, this approach is incongruous with the needs in LTC. The application of a computer-vision based solution in LTC where caregivers and dietitians assess and report residents' food and fluid intake needs has yet to be addressed. Here, the environment is relatively constrained as menu items are relatively set, portion sizes are regulated, and many individuals are monitored by caregivers at each meal.

To address this need, we conducted a participatory iterative design sprint with representative end-users (\eg , personal support workers, dietitians etc.) to inform the needs and problem space, refine the design, and evaluate the resulting early stage user-interface prototype for usability~\cite{pfisterer2019}. User centred and participatory action design methodologies can help overcome barriers to uptake~\cite{leo2017,ienca2018,novitzky2015} and early on user studies and ethical implication consideration of technology are major predictors for user acceptance~\cite{ienca2018}. While the general population may distrust artificial intelligence and related fields including computer vision~\cite{pegasystems2017,pwc2017}, we present a different perspective of positive trust and the need to be responsible with that trust because our designs will be taken at face value. Here, we present one finding that stood out: higher perceived trust in our low-fidelity prototype than in the existing system. Using data from our prototype development, this paper discusses several ethical considerations within the context of applied computer vision techniques and systems in an interdisciplinary environment.

\section{Methods}

A 6-stage iterative participatory design sprint was developed and executed to address the importance and need for user engagement as part of ongoing computer vision driven technology. These six stages included: (1) Design Ideation, (2) Reflect and Storyboard, (3) Storyboard Critiques, (4) Design of the Goldilocks Quality Horizontal Prototype, (5) Usability Assessment, and (6) Final Validation. User feedback and output from each stage was incorporated to inform each proceeding stage. A total of 38 participants and advisors representing 15 distinct roles (\eg personal support worker, nurse, and dietitian) were engaged in the design sprint.  Subjective workload (Raw Task Load Index), subjective usability scales, and a modified Ravden checklist were used to assess project advisors’ perceptions of the AFINI-T system prototype compared with the existing method of food and fluid intake charting as described in~\cite{pfisterer2019}. Here we focus on the output 5 project advisors' evaluation of user perceptions of trust in the AFINI-T system compared to the existing method of food and fluid intake charting. A subset of Jian  \etal~\cite{jian2000}'s tool was used to capture perceptions related to deception, wariness, confidence, dependability, reliability, trust and familiarity with the system~\cite{hoff2015}. Statements were comprised of 7-point Likert scales ranging from not at all (1) to extremely (7). Responses were re-categorized from a 7-point Likert scale to ``No", ``Neutral", and ``Yes" to summarize trends; the original 7-point Likert scale ratings were used for calculating a two-tailed t-test assuming unequal variances~\cite{field2013discovering,norman2008biostatistics} to compare the existing electronic paper-based system and the AFINI-T prototype~\cite{jian2000}.

\section{Results}
Consistent with our previously reported high usability (SUS score of 89.2), and significantly higher perceived performance with the AFINI-T prototype than the existing system (P$<.05$) in~\cite{pfisterer2019}, Table~\ref{tab:title} indicates there is low trust in the existing system (55\% of respondents do not trust the system, n=11), the AFINI-T system was perceived to be more trustworthy, and generally, AFINI-T system trust ratings were opposite and more positive compared to the existing system for food and fluid intake charting (Table~\ref{tab:title}). For example, advisors rated the AFINI-T system as less deceptive (deceptive ``yes”: 17\% AFINI-T, 45\% existing system), less wary of the system (wariness ``yes”: 0\% AFINI-T, 50\% existing system), and more confident in the AFINI-T system (confident ``yes”: 83\% AFINI-T, 18\% existing system). As shown in Table~\ref{tab:title} using a two-tailed t-test assuming unequal variances ~\cite{field2013discovering,norman2008biostatistics} indicate that these three scores of deceptiveness, wariness, and confidence were significantly different between systems all with p-values$<0.05$. While not statistically significant, there were additionally higher ratings for dependability, reliability, and familiarity with the AFINI-T system prototype compared to the existing system.  The statement regarding ``I can trust" the system was higher for AFINI-T system prototype than to the existing system and this difference approached significance (p=0.08).

\begin{table*}
\caption {Advisors' perceived trust of existing food/fluid intake system (Existing) to the AFINI-T prototype (AFINI-T). 7-point Likert scale ratings were condensed to ``No'' for Likert ratings 1-3, ``neutral'' for a rating of 4, and ``yes'' for ratings 5-7.} \label{tab:title}
\begin{tabular}{lllllll}
\hline
\multicolumn{1}{c}{\multirow{2}{*}{\textbf{Trust Statement}}} & \multicolumn{1}{c}{\multirow{2}{*}{\textbf{System}}} & \multicolumn{1}{c}{\multirow{2}{*}{\textbf{\begin{tabular}[c]{@{}c@{}}``No''\\ n (\%)\end{tabular}}}} & \multicolumn{1}{c}{\multirow{2}{*}{\textbf{\begin{tabular}[c]{@{}c@{}}``Neutral''\\ n (\%)\end{tabular}}}} & \multicolumn{1}{c}{\multirow{2}{*}{\textbf{\begin{tabular}[c]{@{}c@{}}``Yes''\\ n (\%)\end{tabular}}}} & \multicolumn{1}{c}{\multirow{2}{*}{\textbf{\begin{tabular}[c]{@{}c@{}}Total\\ n\end{tabular}}}} & \multicolumn{1}{c}{\multirow{2}{*}{\textbf{\begin{tabular}[c]{@{}c@{}}t, df\\ P-value\end{tabular}}}} \\
\multicolumn{1}{c}{} & \multicolumn{1}{c}{} & \multicolumn{1}{c}{} & \multicolumn{1}{c}{} & \multicolumn{1}{c}{} & \multicolumn{1}{c}{} & \multicolumn{1}{c}{} \\ \hline
\multirow{2}{*}{The food/fluid intake system is deceptive.} & Existing & 1 (9\%) & 4 (45\%) & 4 (45\%) & 11 & \multirow{2}{*}{\begin{tabular}[c]{@{}l@{}}3.45, df=11.2\\ (P = .005)\end{tabular}} \\
 & AFINI-T & 5 (83\%) & 0 (0\%) & 1 (17\%) & 6 &  \\ \hline
\multirow{2}{*}{I am wary of the food/fluid intake system.} & Existing & 3 (30\%) & 2 (20\%) & 5 (50\%) & 10 & \multirow{2}{*}{\begin{tabular}[c]{@{}l@{}}3.34, df = 12.5\\ (P = .006)\end{tabular}} \\
 & AFINI-T & 5 (83\%) & 1 (17\%) & 0 (0\%) & 6 &  \\ \hline
\multirow{2}{*}{I am confident in the food/fluid intake system.} & Existing & 5 (45\%) & 4 (36\%) & 2 (18\%) & 11 & \multirow{2}{*}{\begin{tabular}[c]{@{}l@{}}2.42, df = 8.03\\ (P = .042)\end{tabular}} \\
 & AFINI-T & 1 (17\%) & 0 (0\%) & 5 (83\%) & 6 &  \\ \hline
\multirow{2}{*}{The food/fluid intake system is dependable.} & Existing & 4 (40\%) & 3 (30\%) & 3 (30\%) & 10 & \multirow{2}{*}{\begin{tabular}[c]{@{}l@{}}1.55, df = 6.80\\ (P = .166)\end{tabular}} \\
 & AFINI-T & 1 (17\%) & 0 (0\%) & 5 (83\%) & 6 &  \\ \hline
\multirow{2}{*}{The food/fluid intake system is reliable.} & Existing & 5 (50\%) & 2 (20\%) & 3 (30\%) & 10 & \multirow{2}{*}{\begin{tabular}[c]{@{}l@{}}1.07, df = 7.83\\ (P = .319)\end{tabular}} \\
 & AFINI-T & 1 (17\%) & 1 (17\%) & 4 (67\%) & 6 &  \\ \hline
\multirow{2}{*}{I can trust the food/fluid intake system.} & Existing & 5 (45\%) & 4 (36\%) & 2 (18\%) & 11 & \multirow{2}{*}{\begin{tabular}[c]{@{}l@{}}2.00, df = 8.29\\ (P = .080)\end{tabular}} \\
 & AFINI-T & 1 (17\%) & 0 (0\%) & 5 (83\%) & 6 &  \\ \hline
\multirow{2}{*}{I am familiar with the food/fluid intake system.} & Existing & 1 (9\%) & 4 (36\%) & 5 (55\%) & 11 & \multirow{2}{*}{\begin{tabular}[c]{@{}l@{}}1.31, df=14.9\\ (P = .210)\end{tabular}} \\
 & AFINI-T & 0 (0\%) & 0 (0\%) & 6 (100\%) & 6 &  \\ \hline
\end{tabular}
\end{table*}
\section{Discussion and Conclusion}
We were particularly interested in trust as it pertains to the ability to introduce a new level of automation in this field by leveraging computer vision techniques. Several factors described by others that influence trust relevant to developing computer-vision-based applications are: the type of system, how complex the system is, how the system will be used, and the cost-benefit of using the system (\eg , high risk or low risk)~\cite{hoff2015}. However, in the case of the AFINI-T system, we must be careful since this is a tool to support care in a vulnerable setting (older adults). As such, it must be clear that decision making rests with humans and reinforce that the AFINI-T system is a decision-making aid. Specifically, when automation comes with great benefit and fewer risks, people tend to increase reliance on automation; even under high-risk scenarios if the level of automation is low~\cite{hoff2015,lyons2012} and humans tend to view automated tools as more accurate than humans~\cite{lyons2012}. This contributes to over-trust~\cite{lee2004} and may be more important to consider than initially thought given readiness to accept the prototype-level technology and high trust ratings for even for the prototype that is not fully developed.

Our data collection strategy was grounded in theory, guided by several conceptual frameworks, and grounded expertise to complement the interdisciplinary and complexity of the problem space (\eg ,~\cite{boger2017,carr2017}). We also borrowed from transdisciplinary research that ``explicitly recognizes the value of partnerships and the different stakeholders along with their roles in facilitating and supporting innovation''~\cite{boger2017}. In the AFINI-T design this was reflected through recruiting diverse and multidisciplinary project advisors. Our participatory iterative design sprint was employed to work directly with representative end-users to understand and incorporate their perspective and concerns to consider and appropriately support trust from multiple angles. By doing so, we aimed to support trust cues, credibility~\cite{corritore2003,fogg1999}, and to adhere to best practices for user interface design~\cite{NavBars2018,ButtonUX2018,UIdesign2018,nielsen1995,norman2013design,shneiderman2016}. This translated to incorporating ease of navigation (reinforced by click-saving features), use of good visual design elements by on-screen chunking of relevant information, aiming for an overall professional look, supporting search ability as well as smart guidance through transactions and overall, working directly with users to provide appropriate and useful content.

Particularly in computer vision research building intelligent systems with end-users engaged, we should be aware of how the perception of the technology changes through this process as suggested through this work. Specifically, we must be wary of potential over-trust in the system, which is when “trust exceeds system capabilities”~\cite{lee2004}. Of interest in this application, the risk for over-trust is higher when building tools that will be used in high stress environments due to time constraints and have potentially large benefits over the existing method (\eg time saving, improved quality of care)~\cite{hoff2015}. Regarding the receptivity to the AFINI-T system, advisors commented ``Having data accessible opens door to how can be analyzed'' and ``This would save a lot of time especially if individualized. [The \% daily value of nutrients is a] proportional calculation [that is currently] a manual process.'' In contrast, perhaps contributing to relatively low trust in the existing system, several users articulated that the quality of the existing method for data collection is not helpful at prioritizing resident referrals to dietitians. When considered in the LTC environment where being short-staffed is the norm, this may in part explain why trust ratings in the AFINI-T prototype were high and highlights the importance of these design considerations.

While fast-paced and time intensive, the results gathered in this study indicate it \textit{is} possible to design AI/ML/CV applications with end-users and can result in high receptivity and trust in these technologies. Our experience corroborates that engaging with end-users throughout the process as collaborators enhanced a “more comprehensive understanding of the problem space”~\cite{boger2017}. Moving forward, we must re-asses trust in the fully functional system and probe more deeply into factors contributing to trustworthiness. More generally, as designers and researchers developing accountable computer vision systems we must be the first line of defense and should consider our moral obligation for ensuring accuracy, system reliability, and understanding the psychological effects of using trust cues to enhance usability. (Funding: Canada Research Chairs program; NSERC Postgraduate Scholarship-Doctoral)

{\small
\bibliographystyle{ieee}
\bibliography{egbib}
}

\end{document}